\documentclass[final,5p,times,twocolumn,authoryear]{elsarticle}

\usepackage{amssymb}
\usepackage{amsmath}
\usepackage{color} 
\usepackage{comment} 

\usepackage{lineno}

\journal{Ocean Modelling}

\begin{document}

\begin{frontmatter}



\title{Energetically consistent localised APE budgets
for local and regional studies of stratified flow energetics}


\author[inst1]{R\'emi Tailleux}

\affiliation[inst1]{organization={Department of Meteorology, University of Reading},
            addressline={Whiteknights road, Earley Gate}, 
            city={Reading},
            postcode={RG6 6ET}, 
            country={United Kingdom}}

\author[inst2]{Guillaume Roullet}

\affiliation[inst2]{organization={
Université de Bretagne Occidentale, CNRS, IRD, Ifremer, Laboratoire d’Océanographie Physique et Spatiale (LOPS)
},
            addressline={rue Dumont Durville}, 
            city={Plouzané},
            postcode={29280}, 
            country={France}}

\begin{abstract}
Because it allows a rigorous separation between reversible and irreversible
processes, the concept of available potential energy (APE) has become
central to the study of turbulent stratified fluids. In ocean modelling, it
is fundamental to the parameterisation of meso-scale ocean eddies and of the
turbulent mixing of heat and salt. However, how to apply APE theory consistently
to local or regional subdomains has been a longstanding source of confusion due
to the globally defined Lorenz reference state entering the definition of APE
and of buoyancy forces being generally thought to be meaningless in those cases.
In practice, this is often remedied by introducing heuristic `localised' forms of
APE density depending uniquely on region-specific reference states, possibly 
diverging significantly from the global Lorenz reference state. 
In this paper, we argue that across-scale energy
transfers can only be consistently described if 
localised forms of APE density are defined as the eddy APE component of
an exact mean/eddy decomposition of the APE density, for which a
new physically more intuitive and mathematically simpler framework is
proposed. The eddy APE density
thus defined exhibits
a much weaker dependency on the global Lorenz reference state than the mean APE,
in agreement with physical intuition, but with a different structure
than that of existing heuristic localised APE forms. 
Our framework establishes a rigorous physical basis for linking
parameterised energy transfers to molecular viscous and 
diffusive dissipation rates. We illustrate its potential usefulness
by discussing the energetics implications of 
standard advective and diffusive parameterisations of the 
turbulent density flux, which reveals potential new sources of
numerical instability in ocean models.
\end{abstract}

\begin{graphicalabstract}
\includegraphics{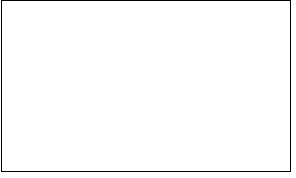}
\end{graphicalabstract}

\begin{highlights}
\item Eddy APE defined through exact mean/eddy APE decomposition rather 
than via heuristic
\item Simpler and more physically intuitive derivation of eddy APE budgets
\item Exact mean-to-eddy APE conversion can be negative, unlike QG
approximation
\item Two new parameters quantify large departures from Lorenz reference
state 
\item Foundation for linking parameterised energy transfers to observable
dissipation rates 
\end{highlights}


\end{frontmatter}



\section{Introduction}

Turbulent stratified flows exhibit a complex interplay between reversible and irreversible processes, each providing crucial insights into the other. Reversible aspects are typically associated with adiabatic stirring, which deforms isopycnal surfaces, increases their areas, and magnifies irreversible effects by enhancing tracer gradients. This process leads to the dissipation of mechanical energy and tracer variances at increasingly smaller scales through molecular and diffusive processes \citep{Eckart1948}. The concept of available potential energy (APE), originally formulated by \citet{Margules1903} and \citet{Lorenz1955} and later adapted to the study of turbulent stratified mixing by \citet{Winters1995}, serves as a key tool for distinguishing between reversible and irreversible processes. APE theory posits that the potential energy (PE) of any stratified fluid can be partitioned into a component (the APE) available for {\em reversible} conversions with kinetic energy (KE), and a component (the background potential energy, BPE) that is not. In Lorenz's approach, the BPE is defined as the PE of a flattened state of minimum potential energy obtainable from the actual state through an adiabatic rearrangement of mass \cite{Lorenz1955}. Consequently, APE theory provides a natural framework for assigning distinct energetic signatures to reversible and irreversible processes \citep{Butler2013}. Reversible processes affect the APE of the fluid while leaving the BPE unaffected, whereas irreversible processes entail an energy transfer between the APE and BPE. In most cases, the net transfer occurs from the APE to the BPE; however, the reverse conversion is occasionally possible, as observed in double-diffusive instabilities (e.g., \citet{Middleton2020,Middleton2021,Tailleux2024}). In the local theory of APE, the conversion rate between APE and BPE is generally referred to as the APE dissipation rate, denoted as $\varepsilon_p$. Although initially introduced in the context of Boussinesq fluids, the nature of $\varepsilon_p$ in the fully compressible Navier-Stokes equations has been discussed by \citet{Tailleux2009,Tailleux2013c,Tailleux2024}. In the global APE framework of \citet{Winters1995}, the volume-integrated APE dissipation rate is represented by the term $\Phi_d - \Phi_i$. Together with the viscous dissipation rate $\varepsilon_k$, the sum $\varepsilon_k+\varepsilon_p$ represents the total dissipation of mechanical energy, defined as the sum of KE and APE.

While recent discussions of APE have primarily been framed within the global APE framework of \citet{Winters1995}, the concept's importance and utility were first recognised by \citet{Oakey1982} and \citet{Gargett1984b} through the derivation of a local APE budget for a quadratic, non-negative APE density. This was achieved by rescaling the budget of temperature variance and linking it to the mechanical energy budget. This result was significant for connecting turbulent mixing to the mechanical energy budget, as it introduced a conversion term with kinetic energy and related the APE dissipation rate $\varepsilon_p$ to the dissipation of temperature variance:
\begin{equation}
        \varepsilon_p =  \frac{g \alpha \kappa_T |\nabla \theta'|^2}{d\overline{\theta}/dz} .
        \label{APE_dissipation_measured} 
\end{equation}
In (\ref{APE_dissipation_measured}), $g$ represents the acceleration due to gravity, $\alpha$ is the thermal expansion coefficient, $\kappa_T$ is the thermal diffusivity, and $\theta$ is the potential temperature with mean $\overline{\theta}$ and perturbation $\theta'$. The relative importance of diffusive and viscous effects in dissipating mechanical energy can be quantified using the dissipation ratio $\Gamma = \varepsilon_p/\varepsilon_k$, a commonly used measure of mixing efficiency, often considered to be close to $0.2$. Both $\varepsilon_p$ and $\varepsilon_k$ can be used to define the turbulent diapycnal mixing according to the formula
\begin{equation}
     K_{\rho} = \frac{\varepsilon_p}{\overline{N}^2}  
     = \frac{\Gamma \varepsilon_k}{\overline{N}^2} ,
\end{equation}
(e.g., \citet{Lindborg2006}). Over the past three decades, the somewhat ad-hoc approach to APE developed by \citet{Oakey1982} and \citet{Gargett1984b} has been superseded by the exact global APE framework of \citet{Winters1995} and the exact finite-amplitude local APE framework first developed by \citet{Andrews1981} and \citet{Holliday1981}, and subsequently extended by \citet{Shepherd1993}, \citet{Scotti2006}, \citet{Roullet2009}, \citet{Tailleux2013b}, \citet{Scotti2014}, \citet{Zemskova2015}, and \citet{Tailleux2018}, among others.

\bigskip 

Reversible and irreversible effects can also be described in terms of energy transfers between different scales of motion. In the atmosphere, there has been extensive discussion about the physical explanation for the turbulent energy cascade affecting both kinetic and potential energy. \citet{Lindborg2006} developed a theory suggesting a forward energy cascade with energy spectra given by
\begin{equation}
    E_{K_h} = C_1 \varepsilon_K^{2/3} k_h^{-5/3}, \qquad
    E_{P_h} = C_2 \varepsilon_p k_h^{-5/3} \varepsilon_K^{-1/3} 
    \label{lindborg_relations}
\end{equation}
with $C_1 \approx C_2$. Interestingly, this theory predicts that the ratio of potential energy to kinetic energy spectra at all scales is 
\begin{equation}
    \frac{E_{P_h}}{E_{K_h}} \approx \frac{\varepsilon_p}{\varepsilon_K} 
    = \Gamma ,
\end{equation}
\textcolor{black}{thus suggesting that the constraint imposed by the dissipation ratio $\Gamma$ extends to significantly larger scales than the dissipation scales. Intriguingly, the ratio $G_A/G_K$ of the APE production rate $G_A$ by surface buoyancy fluxes to the wind power input into geostrophic motions $G_K$ also appears to be close to the dissipation ratio $\Gamma$. Indeed, based on \citet{Zemskova2015}, $G_A \approx 0.5 \,{\rm TW}$, and $G_K \approx 2-3 \,{\rm TW}$, yielding a ratio $G_A/G_K \approx 0.17-0.25$. Strong constraints on the relative importance of the KE and APE energy spectra and energy transfers therefore appear to exist over a wide range of scales. Understanding the nature of such constraints and exploiting them to constrain the parameterisations of subgridscale processes controlling the unresolved KE and APE energy cascades forms the basis for the development of energetically consistent numerical ocean models.} 

\begin{figure}
\begin{center} 
    \includegraphics[width=8cm]{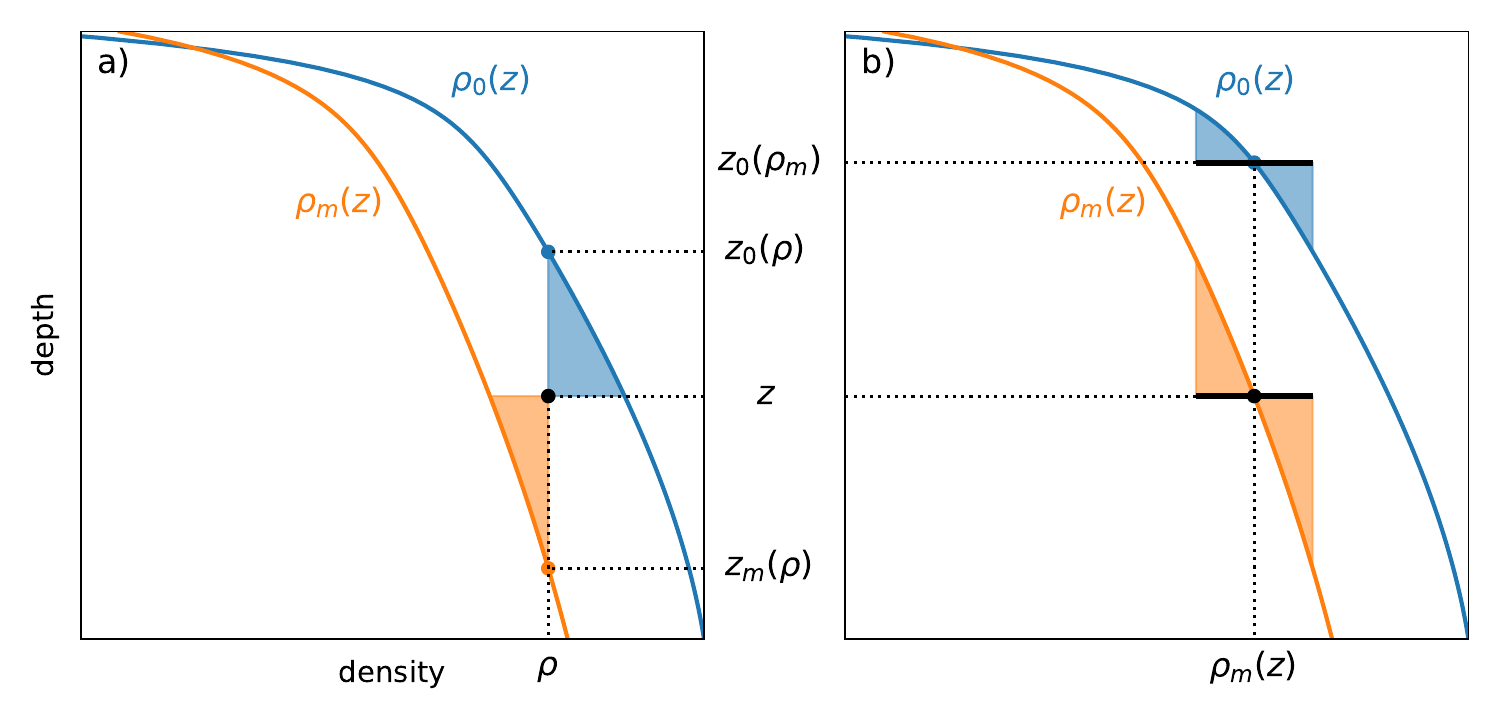}
    \caption{(a) Schematic illustrating the differences between the exact APE density (black) versus heuristic APE density (orange) as area under the curves, for a parcel with density $\rho$ at depth $z$; (b) Same as for left panel but for the instantaneous exact eddy APE density (black) versus heuristic APE density (orange) associated with density anomalies centred around the mean density $\rho_m(z)$ at depth $z$. Note that the black and orange areas can both be approximated as $\rho'\zeta'/2$ in terms of the same density anomaly $\rho'$ but different displacement anomaly $\zeta'$. }
    \label{fig:fig1}
\end{center} 
\end{figure}

Lorenz introduced the global APE framework to explain how the atmospheric circulation is maintained against dissipation. He introduced the Lorenz Energy Cycle (LEC) as a means to achieve this, partitioning the KE and APE reservoirs into mean and eddy components. However, because the integrand defining the APE is not positive definite, Lorenz had to rely on various manipulations and integrations by parts to rewrite the volume-integrated APE as the volume integral of a \textcolor{black}{positive definite} quadratic quantity that could then be split into mean and eddy components. Setting aside compressible effects, the local APE framework defines the APE density of a fluid parcel in terms of the work against buoyancy forces relative to the reference density profile $\rho_0(z)$, characterising the globally defined Lorenz reference state of minimum potential energy. However, this reference state is rarely considered relevant for understanding the energetics of much smaller subdomains. Rather, it is generally implicitly assumed that buoyancy forces should be defined relative to some locally defined averaged density field characteristic of the local environment. To continue using APE theory in such cases, most existing approaches appear to be based on some heuristic `localisation' of the local or global APE frameworks. A typical example of such localisation is the definition of eddy APE by \citet{Roullet2014}, which physically modifies the exact form of local APE density
\begin{equation}
     E_a^{exact} = \frac{g}{\rho_{\star}} \int_{z_0(\rho)}^z 
     [\rho - \rho_0(\tilde{z})]\,{\rm d}\tilde{z} 
     \label{exact_ape_density}
\end{equation}
into the following `localised' form:
\begin{equation} 
     E_a^{R14} = \frac{g}{\rho_{\star}} \int_{z_m}^z
     [\rho - \rho_m(x,y,\tilde{z}) ]\,{\rm d}\tilde{z} ,
     \label{roullet_ape_density} 
\end{equation}
with $\rho_m(x,y,z)$ representing some locally defined mean density field, and $z_m$ the level of neutral buoyancy satisfying $\rho_m(x,y,z_m) = \rho$, 
\textcolor{black}{while $z_0(\rho)$ is such that $\rho_0(z_0(\rho))=\rho$.}
While plausible and physically appealing, (\ref{roullet_ape_density}) is the source of much confusion in the literature about turbulent stratified mixing, as it no longer provides clarity on how to compute the reference state associated with the exact formula (\ref{exact_ape_density}). Several studies have discussed the issue (e.g., \citet{Arthur2017,Davies-Wykes2015,Dewar2019}) and found that different choices of reference states often lead to significantly different conclusions about the properties of mixing. 
\textcolor{black}{To help visualise the nature of the problem, the left panel of Fig. \ref{fig:fig1} illustrates the differences between the exact (black) and heuristic (orange) APE densities predicted by (\ref{exact_ape_density}) and (\ref{roullet_ape_density}) in a particular example for which $\rho_m$ and $\rho_0$ differ from each other, as is normally the case. The right panel anticipates the results of this paper, comparing the exact form of instantaneous eddy APE density resulting from an exact mean/eddy decomposition (black) with the heuristic APE density (orange, same as in left panel), associated with fluctuations around the mean density $\rho_m$. This schematic shows that while both approaches are associated with the same density anomalies $\rho'$, they are associated with different displacement anomalies $\zeta'$, potentially leading to significant differences.} 

Part of the difficulty or confusion surrounding this issue seems to arise from the insistence on discussing the energetics of individual mixing events independently of the energetics of the global ocean in which they are embedded. However, in a statistically steady state, the total KE+APE dissipation must balance the work done by the surface wind stress plus the APE production by the surface buoyancy fluxes. Considering that the APE production by surface buoyancy fluxes is always based on the exact APE density using the Lorenz reference density profile $\rho_0(z)$, it seems evident that if the APE dissipation rate based on (\ref{roullet_ape_density}) is sensitive to the choice of $\rho_m$, then there must be only one consistent way to define $\rho_m$ that can achieve the desired balance. The question addressed in this paper is: which one is it? 

\textcolor{black}{To ensure consistency between APE production at large scales and APE dissipation at molecular scales, the solution discussed in this paper is to define localised eddy APE densities in terms of the eddy component of an exact mean/eddy decomposition of the local APE density. As shown in this paper, this is especially important and necessary where the actual state departs significantly from the Lorenz reference state and hence where the standard QG approximation becomes inaccurate. So far, a general theory for how to do so has been lacking, as existing results have been limited to a Boussinesq fluid with a linear equation of state \citep{Scotti2014} or a dry atmosphere \citep{Novak2018}, using methods too specialised to be easily generalisable. It follows that to apply existing theory to a realistic ocean (e.g., \citet{Zemskova2015} or \citet{MacCready2016}), several approximations need to be made to circumvent the difficulties arising from the two-component nature of the equation of state and its thermobaric nonlinearity, thus diminishing its benefits. Perhaps because of this, most studies of the Lorenz energy cycle --- while often acknowledging the existence of \citet{Scotti2014} --- still continue to rely on the standard QG approximation used by \citet{vonStorch2012} and others. }

\textcolor{black}{In this paper, we revisit \citet{Scotti2014}'s approach by making it more physically intuitive, mathematically simpler, and more easily generalisable to more complex fluids. To that end, we adopt as our starting point the local budget of APE density developed by \citet{Tailleux2024}, which more naturally generalises to two-component fluids. To summarise, our approach defines the mean APE density $E_a(\overline{\rho},z)$ as the APE density of the mean density field $\overline{\rho}$, the eddy APE density as the residual $E_a^t = \overline{E_a} - E_a(\overline{\rho},z)$, and proves the positive definite character of $E_a^t$ by linking it to the convexity of $E_a(\rho,z)$ with respect to $\rho$. The generality of such results suggests that for a realistic thermobaric ocean, the mean APE density should be defined as $E_a^m = E_a (\overline{S}, \overline{\theta},z)$ and hence the eddy APE density as the residual $E_a^t = \overline{E_a} - E_a(\overline{S},\overline{\theta},z)$, which is quite different from \citet{Zemskova2015} or \citet{MacCready2016}, thus paving the way for improving the description of the Lorenz energy cycle in ocean models, as will be developed and discussed in subsequent work.} 

\textcolor{black}{By interpreting the mean and eddy fields as the resolved and unresolved components of a numerical ocean model, we use our framework to discuss and understand the energetic implications of different possible parameterisations of the turbulent fluxes affecting the across-scale KE and APE energy transfers, similarly to the energetically consistent modelling approach of Carsten Eden \citep{Eden2014,Eden2015,Eden2016} (note, though, that in the latter studies, potential energy is partitioned into dynamic and potential enthalpy rather than into APE and BPE). Our main focus is on advective and diffusive parameterisations of the turbulent density flux, which in the oceanic case pertain to meso-scale and small scales, but which in our framework cannot be distinguished. Our discussion reveals that some conclusions about the impact of turbulent parameterisations depend on whether the exact or QG approximation is used; it also suggests that such parameterisations can occasionally convert unresolved APE into resolved APE, potentially revealing a new type of numerical diffusive instability. We also find that if the energy transfers associated with the APE cascade are assumed to scale proportionally to the KE cascade, aspects of the parameterisations of vertical momentum transfer discussed by \citet{Greatbatch1990} can be recovered.} 

\textcolor{black}{In line with classical discussions of the Lorenz energy cycle, our exact mean/eddy decomposition framework relies on conventional Eulerian Reynolds averaging. However, we acknowledge that there is increasing interest in understanding the energetics for the thickness-weighted averaged equations in isentropic or isopycnic coordinates (e.g., \citet{Bleck1985,Aiki2016,Loose2023}). How to extend our results to this case is left to future work.} 

\smallskip 

This paper is organised as follows. Section \ref{convexity_section} highlights the convexity of the local APE density as the fundamental property underlying the construction of the concept of eddy APE in the most general case. The convexity property of the local APE density was only briefly mentioned by \citet{Scotti2014} but arguably warrants a more thorough discussion and exploitation. This section also clarifies the links between exact and heuristic forms of local APE density. Section \ref{budgets_section} revisits and simplifies the derivation of the local budgets of mean and eddy APE previously obtained by \citet{Scotti2014}. Section \ref{constraint_on_mixing} discusses the constraints on mixing parameterisations derived from the consideration of the eddy APE and KE budgets, a key issue for the development of energetically consistent numerical ocean models, which do not appear to have been considered before. Section \ref{summary_and_discussion} provides a summary and discussion of the results.

\section{Convexity and eddy APE density}
\label{convexity_section} 

\subsection{Boussinesq model equations} 

In this study, we analyse the energetics of rotating stratified flows using 
the standard Boussinesq approximation. We define the system's state relative
to the Lorenz reference state, which represents the configuration of minimum
potential energy achievable through an adiabatic rearrangement of fluid parcels.
This reference state is characterised by the pressure and density profiles,
$p_0(z)$ and $\rho_0(z) = -g^{-1} dp_0/dz$, respectively. With these assumptions,
the governing equations of motion may be written as
\begin{equation}
    \frac{D\mathbf{v}}{Dt} + 2 \boldsymbol{\Omega} \times \mathbf{v} + \frac{1}{\rho_{\star}} \nabla p_{\ell}  = b_{\ell} \mathbf{k} + \nu \nabla^2 \mathbf{v}
\label{standard_momentum}
\end{equation}
\begin{equation}
  \nabla \cdot \mathbf{v} = 0 
\end{equation}
\begin{equation} 
     \frac{D\rho}{Dt} = -\nabla \cdot \mathbf{J}_{\rho} , 
     \qquad \mathbf{J}_{\rho} = - \kappa \nabla \rho .
     \label{density_equation} 
\end{equation}
where $p_{\ell} = p-p_0(z)$ is the pressure anomaly relative to the reference
pressure, ${\bf v} = (u,v,w)$ is the three-dimensional velocity field,
$\boldsymbol{\Omega}$ is the rotation vector, $p$ is the pressure, 
$\rho$ is the density, $\kappa$ is the molecular diffusivity, $\nu$ is the kinematic viscosity,
$\rho_{\star}$ is the constant reference Boussinesq density, and $g$ is the acceleration of gravity. 

The buoyancy term in \eqref{standard_momentum} is defined as
\begin{equation}
    b_{\ell} = -\frac{g(\rho-\rho_0(z))}{\rho_{\star}} .
\end{equation}
This buoyancy $b_{\ell}$ differs from the standard buoyancy 
$b_{bou} = -g(\rho-\rho_{\star})/\rho_{\star}$ as it is measured relative to
the variable reference density $\rho_0(z)$ rather than a constant Boussinesq
density $\rho_{\star}$. For the purpose of this analysis, we assume that the
overall domain, analogous to oceanic conditions, is sufficiently large that the
reference density profile $\rho_0(z)$ can be considered time-independent. 
This assumption is supported by climatological observations of temperature 
and salinity over a century (not shown), which indicate that below the mixed
layer, the Lorenz reference state appears to remain stable over time.

\subsection{Local APE theory}

The local APE theory, which builds upon the global APE theory by
\citet{Lorenz1955}, was initially developed by 
\citet{Andrews1981} and \citet{Holliday1981}, and later rooted 
in Hamiltonian theory by \citet{Shepherd1993}. This theory has been further extended and refined by \citet{Scotti2006}, \citet{Roullet2009},
\citet{Scotti2014}, \citet{Zemskova2015}, \citet{Tailleux2013b,Tailleux2018} among others. For a comprehensive review, see \citet{Tailleux2013}. Unlike Lorenz's
global APE theory, the local APE theory defines APE as a local, non-negative
quantity, expressed through an APE density function whose precise form depends on 
the equation of state and the approximations used. 
 
In this paper, we specifically consider a standard Boussinesq fluid with 
a linear equation of state. The expression for the APE density, derived from
\citet{Holliday1981} and subsequently utilised by \citet{Roullet2009} and
\citet{Tailleux2013b}, is: 
\begin{equation}
     E_a(\rho,z)
     = \frac{g}{\rho_{\star}} \int_{z_0(\rho)}^z [\rho-\rho_0(\tilde{z}) ]\,{\rm d}\tilde{z} = - \int_{z_0(\rho)}^z b_{\ell} (\rho,\tilde{z})
     \,{\rm d}\tilde{z} 
     \label{APE_density} 
\end{equation}
where $z_0(\rho)$ is the Level of Neutral Buoyancy (LNB) at which the density of
a fluid parcel equals that of the reference density: 
\begin{equation}
      \rho = \rho_0 (z_0(\rho)) .
      \label{lnb_equation} 
\end{equation}
Note that in Equation \eqref{APE_density}, $\rho-\rho_0(\tilde{z})$ should be
interpreted as $\rho(x,y,z,t)-\rho_0(\tilde{z})$, with $\rho$ held constant
during the integration. The APE density, like other forms of exergy  
\citep{Marquet1991,Kucharski1997,Kucharski2001}, 
is an extrinsic state function, dependent
on both the fluid parcel's state and its environmental context. 

The APE density can also be expressed in terms of $\rho$:
\begin{equation}
    E_a(\rho,z) = 
          \int_{\rho^{\ddag}(z)}^{\rho} 
          \frac{\partial E_a}{\partial \rho}(\tilde{\rho},z) \,{\rm d}\tilde{\rho}
   =  \frac{g}{\rho_{\star}} \int_{\rho^{\ddag}(z)}^{\rho}  
    [ z-z_0(\tilde{\rho}) ] \,{\rm d}\tilde{\rho} ,
    \label{APE_density_rho_coordinates}
\end{equation}
where $\rho^{\ddag}$ is defined such that $z_0(\rho^{\ddag}) = z$. 
Equation \eqref{APE_density_rho_coordinates} is the starting point of the
mean/eddy decomposition obtained by \citet{Scotti2014} and is often favoured
over (\ref{APE_density}) in the literature. In this study, 
Eq. (\ref{APE_density}) is preferred over (\ref{APE_density_rho_coordinates}),
as it aligns more closely with the APE density for a multi-component 
compressible fluid, e.g., \citet{Tailleux2013b,Tailleux2018}.

Given that $\rho$ is a function of position and time, $E_a(\rho,z)$ 
can also be viewed as a function of $(x,y,z,t)$. To differentiate between
vertical derivatives calculated at constant $(x,y,t)$ versus at constant $\rho$,
we introduce the two separate notations:
\begin{equation}
    \frac{\partial}{\partial z} = \left .
    \frac{\partial}{\partial z} \right |_{x,y,t}   \qquad {\rm versus}
    \qquad 
    \frac{\partial}{\partial Z} = \left .
    \frac{\partial}{\partial z} \right |_{\rho} .
\end{equation}
The partial derivatives of $E_a$ with respect
to density and height are: 
\begin{equation}
     \frac{\partial E_a}{\partial \rho} 
     = \frac{g (z-z_0(\rho))}{\rho_{\star}} = \frac{g\zeta}{\rho_{\star}} = 
     \Upsilon 
     \label{rho_derivative}
\end{equation}
\begin{equation}
     \frac{\partial E_a}{\partial Z} = \frac{g(\rho-\rho_0(z))}{\rho_{\star}} 
     = - b_{\ell} .
     \label{z_derivative} 
\end{equation}
Here, $\zeta$ represents the displacement from the reference depth $z_0(\rho)$,
and $\Upsilon$ denotes a thermodynamic efficiency factor, indicating how 
diabatic heating influences APE density versus background potential energy (BPE).
This concept aligns with the thermodynamic efficiency in compressible fluids,
as discussed by \citet{Tailleux2024}.

The functional dependence of $E_a$ on $\rho$
and $z$ leads to the following identity
\begin{equation}
    \nabla E_a = \frac{\partial E_a}{\partial \rho} \nabla \rho 
    + \frac{\partial E_a}{\partial Z} \nabla z 
    = {\bf P}_a - b_{\ell} {\bf k}
    \label{crocco_theorem} 
\end{equation}
in which
\begin{equation}
{\bf P}_a = \frac{\partial E_a}{\partial \rho} \nabla \rho = \Upsilon \nabla \rho .
\end{equation}
This equality represents a particular instance of the Crocco-Vazsonyi theorem 
\citep{Crocco1937,Vazsonyi1945}, which is crucial for
developing energetically consistent sound-proof approximations, as recently
explored by \citet{TailleuxDubos2024}. Here, ${\bf P}_a$ denotes an APE-based modification 
of the P-vector originally introduced by \citet{Nycander2011} and recently showed by 
\citet{Tailleux2023} to relate to the directions of lateral 
stirring in the oceans.

\subsection{`Heuristic' eddy APE}

In most studies, buoyancy forces are often introduced
in an ad-hoc manner, reliant on an arbitrary selection
of a reference state \citep{Thorpe1989,Smith2005}. 
These forces, while useful, typically lack intrinsic
physical significance. However, the work done against
the buoyancy forces defined relative to the Lorenz 
reference density profile represents the energy required to achieve the stratification of the 
actual state from the Lorenz reference state through
an adiabatic rearrangement of fluid parcels. This
process has intrinsic dynamical significance since it
is inherently tied to the system's physical properties.
The associated squared buoyancy frequency profile 
for such forces is 
\begin{equation}
    N_0^2(z) = -\frac{g}{\rho_{\star}} \frac{d\rho_0}{dz}(z) .
    \label{reference_buoyancy_frequency} 
\end{equation}
However, it is generally considered that this definition of buoyancy forces pertains primarily to the energetics of the large-scale flows, with less relevance to local turbulent mixing events. For the latter, it is generally assumed that the relevant buoyancy forces are those defined in terms of the buoyancy anomaly: 
\begin{equation}
     b' = - \frac{g}{\rho_{\star}}
     (\rho - \overline{\rho}(x,y,z)) = - \frac{g\rho'}{\rho_{\star}} .
     \label{buoyancy_anomaly} 
\end{equation}
Based on this, an intuitive extension of the APE
density to measure the work against these small-scale
forces consists in adapting (\ref{APE_density}) 
as follows: 
\begin{equation}
       E_a^{heu} = \frac{g}{\rho_{\star}} \int_{z_m(x,y,\rho)}^z 
       [\rho - \overline{\rho} (x,y,\tilde{z}) ]\,{\rm d}\tilde{z} ,
       \label{APE_heuristic}
\end{equation}
Here, $\overline{\rho}$ replaces $\rho_0(z)$, 
and $z_m = z_m(x,y,\rho)$ is the height at which a
fluid has zero buoyancy relative to $\overline{\rho}$, solving 
\begin{equation} 
   \overline{\rho}[x,y,z_m(x,y,\rho)] = \rho .
   \label{LNB_heuristic} 
\end{equation} 
This heuristic APE density underpins the work of 
\citet{Roullet2014} on mesoscale eddy APE from 
ARGO float data, and \citet{Luecke2017}'s comparison of simulated and observed eddy APE. 
At leading order, this
can be approximated by 
\begin{equation}
      E_a^{heu} \approx - \frac{1}{2} b'\zeta' \approx 
     \frac{g^2}{\rho_{\star}^2 \overline{N}^2} \frac{\rho'^2}{2} 
     = \frac{1}{2} \frac{b'^2}{\overline{N}^2} 
\end{equation}
where $\zeta' = z - z_m(x,y,\rho)$ and 
$\overline{N}^2$ is the local mean squared buoyancy
frequency 
\begin{equation}
     \overline{N}^2 = -\frac{g}{\rho_{\star}} \frac{\partial \overline{\rho}}{\partial z} (x,y,z_m) .
\end{equation}

However, defining the APE density of small or mesoscale motions as per (\ref{APE_heuristic}) has issues: 

\medskip 
\par \noindent 
{\bf Energy budget consistency}: 
the evolution equation
for $E_a^{heu}$ includes energy conversion terms 
(such as one proportional to the horizontal density gradient $\nabla_h \overline{\rho}$ for instance) with no counterpart in the eddy kinetic energy equation, complicating its energetics description. 

\medskip 
\par \noindent 
{\bf Link to large-scale APE} 
Physically, the APE dissipation by small scale mixing must ultimately balance, if only partially, 
the large-scale sources of APE imparted at the boundaries of the domain \citep{Zemskova2015}.
Since (\ref{APE_heuristic}) does not depend on $\rho_0(z)$, it is challenging to link 
small-scale turbulent dissipation to large scale 
APE sources.

\medskip 
In this paper, we argue that in order to retain a connection with both $b'$ and $\rho_0(z)$, the eddy APE density needs to be defined as part of an exact mean/eddy decomposition of the APE density. At leading order, this exact eddy APE density is also equal to $-1/2 b'\zeta'$, but with the displacement $\zeta'$ defined as $\zeta' = z_0(\rho) - z_0(\overline{\rho})$. Importantly, our exact eddy APE density is valid for arbitrarily large deviations from Lorenz reference state. As established further in the text, this theory modifies the classical relationship between turbulent diapycnal diffusivity and eddy APE dissipation $\varepsilon_p^t$:
\begin{equation}
     K_{\rho} = \frac{\varepsilon_p^t}{\overline{N}^2} 
     \quad (old) \qquad 
     \Longrightarrow \qquad 
     K_{\rho} = \frac{1}{\Lambda (1+|{\bf S}|^2)} 
     \frac{\varepsilon_p^t}{\overline{N}^2}  \quad (new)
\end{equation}
Here, $\Lambda$ and ${\bf S}$ are 
parameters defined by 
\par \medskip 
\fbox{
\begin{minipage}{8cm} 
\begin{equation}
    \Lambda = \frac{\partial \overline{\rho}}{\partial z} 
    \frac{\partial z_0}{\partial \rho}(\overline{\rho}) 
    = \frac{\overline{N}^2}{\overline{N}_0^2} , \qquad 
{\bf S} = - \left ( 
\frac{\partial \overline{\rho}}{\partial z}
\right )^{-1} \nabla_h \overline{\rho} 
\label{n2_density_slope} 
\end{equation}
\end{minipage}
} 
\medskip 
\par \noindent 
where $\overline{N}_0^2 = N_0^2 (z_0(\overline{\rho}))$. The two parameters $\Lambda$ and ${\bf S}$ measure deviations from Lorenz reference state in different ways and play a crucial role in this paper, as clarified further in the text.
Neglecting $|{\bf S}|$ in the expression for $K_{\rho}$ is equivalent to making the small slope approximation. Physically, $\Lambda$ is expected to differ significantly from unity where the local squared buoyancy differs significantly from its value in Lorenz reference state; in the oceans, this is primarily the case in the polar oceans \citep{Saenz2015,Tailleux2016b,Tailleux2023}. 

\subsection{Convexity of APE density and eddy APE} 

We now separate variables into mean and eddy components using standard 
Reynolds averaging so that ${\bf v} = \overline{\bf v} + {\bf v}'$, $\rho = \overline{\rho}+ \rho'$, and so on, such that for any variable $f$, 
$\overline{\overline{f}} = \overline{f}$ and $\overline{f'}=0$.
As is well known, such an approach yields to the following mean/eddy partition
of the Reynolds averaged kinetic energy
\begin{equation}
     \overline{\frac{{\bf v}^2}{2}} = \frac{\overline{\bf v}^2}{2} 
     + \frac{\overline{{\bf v}'^2}}{2} = E_k^m + E_k^t .
\end{equation}
Importantly, the mean kinetic energy $E_k^m$ appears as the kinetic energy of the
mean velocity field $\overline{\bf v}$. The corresponding problem for the APE 
density is to achieve the following mean/eddy decomposition of the Reynolds 
averaged APE density
\begin{equation}
    \overline{E}_a = E_a^m + E_a^t . 
\end{equation}
Ideally, one would like to define the mean APE density as $E_a^m = E_a(\overline{\rho},z)$, that is, as the APE density of the mean density $\overline{\rho}$,
since the latter is the quantity that appears in the Reynolds averaged momentum
equations, and therefore define the eddy APE as the residual
$E_a^t = \overline{E}_a - E_a(\overline{\rho},z)$. That such an approach leads
to a non-negative $E_a^t$ was established by \citet{Scotti2014}, who linked 
the result to the convexity of APE density with respect to buoyancy but without elaborating on it. However, given its fundamental importance in available energy theories, the convexity property warrants more emphasis.

Mathematically, a function $f(x)$ is said to be convex at some point $x_0$ if its curve lies above its tangent line at that point, hence if the quantity
$f_e(x;x_0) = f(x)-f(x_0) - f'(x_0)(x-x_0) \ge 0$ is non-negative. To clarify 
what properties of $f$ determine its convexity, 
it is useful to rewrite $f_e$ in the
form
\begin{equation}
    f_e = \int_{x_0}^x [f'(\tilde{x}) - f'(x_0)] \,{\rm d}\tilde{x} 
    = \int_{x_0}^x \int_{x_0}^{\tilde{x}} f''(\hat{x}) \,{\rm d}\hat{x}
    {\rm d}\tilde{x} .
    \label{identity_convexity}
\end{equation}
Eq. (\ref{identity_convexity}) is an important identity as it shows that $f$ is 
convex at the point $x_0$ if its second derivative $f''$ is non-negative 
(assuming $f$ to be twice differentiable). Of course, convexity extends to functions
of several variables. For instance, for a function $f(x,y)$, convexity requires
that the quantity $f_e(x,y;x_0,y_0) = f(x,y)-f(x_0,y_0) - \partial_x f (x_0,y_0)(x-x_0) - \partial_y f (x_0,y_0) (y-y_0) \ge 0$ be non-negative. Convexity plays a key
role in thermodynamics. For instance, it can be shown that the possibility to 
convert heat into work, the central object of thermodynamics, hinges on internal
energy being a convex function of its canonical variables,
specific entropy and specific volume, which is key to
defining the concept of exergy, e.g., \citet{Tailleux2013}. Arguably, it is the 
convexity of kinetic energy that is implicitly
responsible for the non-negative character of
eddy kinetic energy. For quadratic expressions, however, there is no need to 
invoke convexity as this is not required to prove the non-negative character of 
the eddy component. Convexity is needed here, however, because the APE density 
includes higher-order non-quadratic terms, called anharmonic by \citet{Roullet2009}.

As it turns out, the APE density (\ref{APE_density}) is convex 
with respect to both density and $z$, which can be verified by differentiating
(\ref{rho_derivative}) and (\ref{z_derivative}) with respect to $\rho$ and $z$
respectively, which leads to
\begin{equation}
     \frac{\partial^2 E_a}{\partial \rho^2} 
     = - \frac{g}{\rho_{\star}} \frac{dz_0}{d\rho} (\rho) ,
     \label{rho_second_derivative}
\end{equation}
\begin{equation}
    \frac{\partial^2 E_a}{\partial Z^2} = -\frac{g}{\rho_{\star}} 
    \frac{d\rho_0}{dz} (z) = N_0^2 (z) . 
    \label{z_second_derivative} 
\end{equation}
Physically, the non-negative character of $N_0^2$ in Eq. \ref{z_derivative}) 
follows from Lorenz reference state being a state of minimum potential energy,
hence statically stable by construction, which establishes convexity with respect
to $z$. To prove the non-negative character of (\ref{rho_derivative}) and hence
the convexity with respect to density, simply differentiate the 
LNB equation (\ref{lnb_equation}) with respect to $\rho$, thus leading to
\begin{equation}
    \frac{d\rho_0}{dz} (z_0(\rho)) \frac{dz_0}{d\rho} (\rho) = 1 .
    \label{useful_identity}
\end{equation}
Now, since $d\rho_0/dz\le 0$, (\ref{useful_identity}) implies that 
$dz_0/d\rho \le 0$, which proves our proposition. 

Having established the convexity of $E_a$ with respect to both density and $z$,
let us define the instantaneous eddy APE as the non-negative 
nonlinear term $A_e$ in perturbation
density in the following series expansion of $E_a$ around $\overline{\rho}$ 
\begin{equation}
   E_a(\rho,z) = E_a(\overline{\rho},z) + 
   \frac{\partial E_a}{\partial \rho}(\overline{\rho},z) (\rho - \overline{\rho}) 
   + A_e .
   \label{ape_decomposition} 
\end{equation}
Upon Reynolds averaging, the term proportional to $\rho' = \rho -\overline{\rho}$ vanishes, leading the mean/eddy decomposition: 
\begin{equation}
     \overline{E_a} = E_a^m + E_a^t, 
\end{equation}
with
\begin{equation}
     E_a^m = E_a (\overline{\rho}, z)  , \qquad
     E_a^t = \overline{A_e} .
\end{equation}
Physically, the instantaneous eddy APE $A_e$ is the counterpart of the instantaneous
value of eddy kinetic energy ${\bf v}'^2/2$. 
To facilitate the comparison of $A_e$ with the heuristic APE density discussed
earlier, it is useful to rewrite $A_e$ in the following more revealing form
\par \medskip 
\fbox{
\begin{minipage}{8cm} 
\begin{equation}
\begin{split} 
     A_e = &  E_a(\rho,z) - E_a(\overline{\rho},z) 
     - \frac{\partial E_a}{\partial \rho} (\overline{\rho},z) 
     (\rho - \overline{\rho} ) \\ 
     = & \int_{\overline{\rho}}^{\rho} 
     \left [ \frac{\partial E_a}{\partial \rho}(\tilde{\rho},z) 
     - \frac{\partial E_a}{\partial \rho}(\overline{\rho},z) \right ] \,{\rm d}\tilde{\rho}  \\
     = & - \frac{g}{\rho_{\star}} \int_{\overline{\rho}}^{\rho} 
     [ z_0 (\tilde{\rho})  - z_0 (\overline{\rho} ) ]\,{\rm d}\tilde{\rho}   
     \label{ae_in_terms_or_rho} 
\end{split} 
\end{equation}
\end{minipage}
}
\par \medskip \noindent 
where the passage from the penultimate to last equation made use of (\ref{rho_derivative}). 
Alternatively, $A_e$ may be rewritten as an integral of the work against buoyancy forces by 
introducing the change of variable $\tilde{\rho} = \rho_0(\tilde{z})$ so that $d\tilde{\rho} 
= \rho_0'(\tilde{z}) \,{\rm d}\tilde{z} $, and 
\par \medskip 
\fbox{
\begin{minipage}{8cm}
\begin{equation}
     A_e = - \frac{g}{\rho_{\star}} \int_{z_0(\overline{\rho})}^
     {z_0(\rho)} [ \rho - \rho_0(\tilde{z}) ] \,{\rm d}\tilde{z} 
     \label{ae_in_terms_of_z} 
\end{equation}
\end{minipage}
}
\par \medskip \noindent 
Importantly, note that (\ref{ae_in_terms_or_rho}) and (\ref{ae_in_terms_of_z}) reveal that $A_e= A_e(\rho,\overline{\rho})$ no longer depends on height $z$, being 
solely a function of $\rho$ and $\overline{\rho}$ only.
In both cases, the expression for the transient eddy APE may be approximated as
\par \medskip 
\fbox{
\begin{minipage}{8cm} 
\begin{equation}
     A_e \approx -\frac{1}{2} b'\zeta' \qquad 
     \Rightarrow \qquad E_a^t \approx - \frac{1}{2} \overline{b'\zeta'} 
     \label{transient_eddy_APE_exact} 
\end{equation}
\end{minipage}
}
\par \medskip \noindent 
with
\begin{equation}
     b' = -\frac{g(\rho-\overline{\rho})}{\rho_{\star}} , 
     \qquad \zeta' = z_0(\rho) - z_0(\overline{\rho}) 
\end{equation}
The approximation (\ref{transient_eddy_APE_exact}) is identical to the expression used
by \citet{Roullet2014}. However, while the buoyancy anomaly $b'$ is the same
as in Roullet's approach, the displacement $\zeta'$ is defined quite differently 
in terms of the instantaneous and mean reference positions
$z_0(\rho)$ and $z_0(\overline{\rho})$ respectively, that is, in terms of the
equilibrium positions of $\rho$ and $\overline{\rho}$ in Lorenz reference state. \textcolor{black}{These differences are clearly
illustrated in the right panel of Fig. \ref{fig:fig1} 
introduced earlier.}

Physically, the approximation (\ref{transient_eddy_APE_exact}) is obtained from 
a simple trapezoidal approximation of the integrals (\ref{ae_in_terms_or_rho})
or (\ref{ae_in_terms_of_z}), and are likely to be the most accurate approximation of $A_e$. Nevertheless, using the approximation
\begin{equation}
    \zeta' = z_0(\rho) - z_0(\overline{\rho}) \approx 
    \frac{\partial z_0}{\partial \rho}(\overline{\rho}) 
    \rho' 
\end{equation}
it is also possible to approximate $A_e$ in terms of the following quadratic
expressions: 
\begin{equation}
    A_e \approx -\frac{g}{\rho_{\star}} 
    \frac{dz_0}{d\rho} (\overline{\rho}) 
    \frac{\rho'^2}{2} = \frac{g^2}{\rho_{\star}^2 \overline{N}_0^2 }
    \frac{\rho'^2}{2} = \frac{1}{2} \frac{b'^2}{\overline{N}_0^2} 
\end{equation}
with
\begin{equation} 
    \overline{N}_0^2 = \overline{N}_0^2(x,y,z) = 
    - \frac{g}{\rho_{\star}} \frac{d\rho_0}{dz} (z_0(\overline{\rho})) .
    \label{correct_mean_n2} 
\end{equation}
\par \medskip \noindent 
Note that $\overline{N}_0^2$, unlike $N_0^2(z)$, is a function of all three spatial
dimensions $(x,y,z)$. Its spatial gradient is easily verified to be
\begin{equation}
    \nabla \overline{N}_0^2 = - \frac{g}{\rho_{\star}} 
    \frac{d^2 \rho_0}{dz^2} \frac{\partial z_0}{\partial \rho} 
    \nabla \overline{\rho} 
\end{equation}
so appears to be proportional to the mean density gradient $\nabla \overline{\rho}$,
the proportionality factor being controlled by the curvature of $\rho_0(z)$. 
Comparison with the heuristic localised APE density $E_a^{heu}$ and $A_e$ is
easily verified to be 
\begin{equation}
    \frac{A_e}{E_a^{heu}} \approx \frac{\overline{N}^2}{\overline{N}_0^2} = 
    \Lambda ,
\end{equation}
where $\Lambda$ is the same parameter introduced previously. This establishes that the validity and accuracy of the heuristic localised APE density depend on the proximity of the actual state to Lorenz reference state, \textcolor{black}{as expected}.

\section{Local budgets of available potential energy} 
\label{budgets_section} 

\subsection{Non-averaged local APE budgets} 

Before deriving local budget equations for the mean and APE densities $E_a^m$
and $E_a^t$, we first clarify the local budget equation satisfied by the non-averaged APE density $E_a$. This can be obtained by taking  the Lagrangian derivative of $E_a$, yielding
\begin{equation}
    \frac{DE_a}{Dt} =
    \left ( \frac{\partial E_a}{\partial \rho} \frac{D\rho}{Dt} 
    + \frac{\partial E_a}{\partial Z} \frac{Dz}{Dt} \right ) 
    = \Upsilon \frac{D\rho}{Dt} - b_{\ell} w .
    \label{lagrangian_ape_evolution} 
\end{equation}
By making use of the density equation (\ref{density_equation}), it is easily checked that
(\ref{lagrangian_ape_evolution}) may be rewritten in the form
\par \bigskip 
\fbox{
\begin{minipage}{8cm} 
\begin{equation}
     \frac{DE_a}{Dt} = - b_{\ell} w
    - \nabla \cdot {\bf J}_a - \varepsilon_p 
    \label{ape_density_evolution_first_form}
\end{equation}
\end{minipage} 
} 
\par 
\medskip 
\noindent
in which ${\bf J}_a$ and $\varepsilon_p$ are the diffusive flux of APE density and
APE dissipation rate, respectively, given by 
\par \medskip
\fbox{
\begin{minipage}{8cm}
\begin{equation}
    {\bf J}_a = \Upsilon {\bf J}_{\rho} = -\Upsilon \kappa \nabla \rho  ,  
\end{equation}
\begin{equation}
    \varepsilon_p = - {\bf J}_{\rho} \cdot \nabla \Upsilon 
    = \kappa \nabla \rho \cdot \nabla \Upsilon 
    \label{ape_dissipation_rate_non_averaged}
\end{equation}
\end{minipage}
}
\par \medskip \noindent 
The normal component of ${\bf J}_a$ at the ocean surface determines the APE
production rate by surface buoyancy fluxes, see \citet{Zemskova2015} for 
a discussion within the present framework. Our evolution equation for the APE density (\ref{ape_density_evolution_first_form}), although mathematically equivalent, is much simpler in form that the one previously derived \citet{Scotti2014} due to not imposing ${\bf J}_a$ to be downgradient in $E_a$. Physically, the form  (\ref{ape_density_evolution_first_form}) is to be preferred because it it the one that most naturally generalise to double diffusive multi-component compressible stratified fluids, unlike \citet{Scotti2014}'s approach, e.g. see \citet{Tailleux2024} for details. Using the expression
(\ref{rho_derivative}) for $\Upsilon$, $\varepsilon_p$ may also be expressed in the more familiar form
\par \medskip 
\fbox{
\begin{minipage}{8cm} 
\begin{equation}
     \varepsilon_p = \frac{g \kappa}{\rho_{\star}} 
     \nabla \rho \cdot \left [ {\bf k} - \frac{dz_0}{d\rho} \nabla \rho \right ] 
     = \frac{g\kappa}{\rho_{\star}}  
     \left [ \frac{\partial \rho}{\partial z} - 
     \frac{dz_0}{d\rho} |\nabla \rho|^2 \right ]  
     \label{ape_dissipation_rate_formula}
\end{equation}
\end{minipage}
}
\par \medskip \noindent 
How the APE density budget equation (\ref{ape_density_evolution_first_form}) relates to that of a fully compressible fluid as well as to the global evolution equations derived by \citet{Winters1995}, is 
extensively discussed by \citet{Tailleux2024}, to which the reader is referred to for details. For the reader more familiar with the global APE approach of \citet{Winters1995}, it may be useful to point out that the volume integral of (\ref{ape_dissipation_rate_formula}) coincides with the 
term $\Phi_d-\Phi_i$ of \citet{Winters1995}. 

\subsection{Mean APE budget} 

As established previously, the mean APE density $E_a^m = E_a(\overline{\rho},z)$ is naturally defined as the APE density of the mean density $\overline{\rho}$ and $z$, at least when approached from the viewpoint of standard Eulerian averaging. An evolution equation for it can therefore be derived essentially as that for the non-averaged density $E_a(\rho,t)$, with $D\rho/Dt$ replaced by $D_m\overline{\rho}/Dt$, the Lagrangian derivative of $\overline{\rho}$ defined in terms of the mean velocity $\overline{\bf v}$, which leads to
\begin{equation}
     \frac{D_m E_a^m}{Dt} 
     = \frac{\partial E_a^m}{\partial \overline{\rho}} 
     \frac{D_m \overline{\rho}}{\partial t} 
+ \frac{\partial E_a^m}{\partial Z} \frac{D_m z}{Dt} 
     = \Upsilon_m \frac{D_m \overline{\rho}}{Dt} 
     - \overline{b}_{\ell} \overline{w} ,
     \label{ddt_eam} 
\end{equation}
in which $\Upsilon_m$ and $\overline{b}_{\ell}$ are defined by
\begin{equation}
    \Upsilon_m = \frac{\partial E_a}{\partial \rho} (\overline{\rho},z) 
    = \frac{g ( z -z_0(\overline{\rho}))}{\rho_{\star}}  .
    \label{upsilon_m} 
\end{equation}
\begin{equation}
    \overline{b}_{\ell} = \frac{\partial E_a}{\partial Z}(\overline{\rho},z)
    = \frac{g (\overline{\rho} - \rho_0(z))}{\rho_{\star}} .
\end{equation}
It is important to note here that the mean quantity $\Upsilon_m$ differs from the Reynolds averaged $\Upsilon$, i.e., 
$\Upsilon_m \ne \overline{\Upsilon}$, because  
$\overline{z_0(\rho)} \ne z_0(\overline{\rho})$ in general. Using the evolution equation for the mean density $\overline{\rho}$, viz., 
\begin{equation}
    \frac{D_m \overline{\rho}}{Dt} = -\nabla \cdot 
    [ \overline{\rho'{\bf v}'} - \kappa \nabla \overline{\rho} ] 
    = - \nabla \cdot {\bf J}_{\rho}^m ,
    \label{mean_density_equation}
\end{equation}
in which ${\bf J}_{\rho}^m = \overline{\rho'{\bf v'}} - \kappa \nabla \overline{\rho}$ is the total density flux including both turbulent and molecular diffusive contributions, it is straightforward to show that (\ref{ddt_eam}) may be rewritten as
\begin{equation}
    \frac{D_m E_a^m}{Dt} 
    = - \overline{b}_{\ell} \overline{w}
    - \nabla \cdot {\bf J}_a^m  
    + {\bf J}_{\rho}^m \cdot \nabla \Upsilon_m 
    \label{ddt_eam_bis}
\end{equation}
where ${\bf J}_a^m$ is the total `diffusive' flux of mean APE density including
both turbulent and molecular contributions, given by 
\begin{equation}
     {\bf J}_a^m = \Upsilon_m {\bf J}_{\rho}^m 
     = \Upsilon_m ( \overline{\rho'{\bf v}'} - \kappa \nabla \overline{\rho} )  
\end{equation}
Note that the last term in (\ref{ddt_eam_bis}) may be further expanded in the
form
\begin{equation}
    {\bf J}_{\rho}^m \cdot \nabla \Upsilon_m 
    = \overline{\rho'{\bf v}'} \cdot \nabla \Upsilon_m 
    - \kappa \nabla \overline{\rho} \cdot \nabla \Upsilon_m 
    = \overline{\rho'{\bf v}'} \cdot \nabla \Upsilon_m - \varepsilon_p^m 
\end{equation}
where 
\begin{equation}
    \varepsilon_p^m = \kappa \nabla \overline{\rho}\cdot \nabla \Upsilon_m
    \label{mean_ape_dissipation_rate} 
\end{equation} 
represents the `mean' APE dissipation rate. As a result, (\ref{ddt_eam_bis}) 
may ultimately be rewritten in the following final form:
\par \medskip
\fbox{
\begin{minipage}{8cm} 
\begin{equation}
    \frac{D_m E_a^m}{Dt} 
    = - \overline{b}_{\ell} \overline{w} 
    + \overline{\rho'{\bf v}'}\cdot \nabla \Upsilon_m 
    - \nabla \cdot {\bf J}_a^m - \varepsilon_p^m 
    \label{ddt_eam_final} 
\end{equation}
\end{minipage}
} 
\par \medskip 
Physically, the terms appearing in the r.h.s. of (\ref{ddt_eam_final}) represent:
1) the conversion between mean APE and mean KE; 2) the conversion between mean
APE and eddy APE; 3) the diffusive flux of mean APE by means of turbulent and 
molecular processes; 4) the mean dissipation rate of APE by molecular processes. 

\subsection{Eddy APE budget and turbulent APE dissipation} 

We now turn to the problem of deriving an evolution equation for the eddy APE $E_a^t = \overline{A_e}$. There are two main routes. The first route is via deriving an 
evolution equation for $A_e$ and Reynolds averaging the result. In the second route, which is much simpler and the only one pursued here, the evolution equation for $E_a^t = \overline{E}_a - E_a^m$ is obtained as 
the residual between the evolution equations for $\overline{E}_a$ and $E_a^m$. 

To proceed, we first take the Reynolds average of  (\ref{ape_density_evolution_first_form}) after separating each variable into mean and eddy components, which leads to
\begin{equation}
    \frac{D_m \overline{E_a}}{Dt} + 
    \nabla \cdot ( \overline{E_a'{\bf v}'} ) 
    = - \overline{b_{\ell}} \overline{w} - \overline{b_{\ell}'w'} 
    - \nabla \cdot \overline{{\bf J}_a} - \overline{\varepsilon_p} 
    \label{Mean_total_APE_evolution} 
\end{equation}
The sought-for evolution equation for $E_a^t$ is then simply obtained by 
subtracting the $E_a^m$ equation (\ref{ddt_eam_final}) from 
(\ref{Mean_total_APE_evolution}), which yields 
\par \medskip 
\fbox{
\begin{minipage}{8cm} 
\begin{equation}
    \frac{D_m E_a^t}{Dt} 
    = - \overline{b_{\ell}'w'} - \overline{\rho'{\bf v}'} \cdot \nabla \Upsilon_m 
    - \nabla \cdot {\bf J}_a^t - \varepsilon_p^t 
    \label{eddy_ape_evolution_first_form}
\end{equation}
\end{minipage} 
} 
\par \medskip \noindent 
in which ${\bf J}_a^t$ and $\varepsilon_p^t$ are the total flux of eddy APE density
and eddy APE dissipation rate respectively, whose expressions are
\begin{equation}
    {\bf J}_a^t = \overline{{\bf J}_a} - {\bf J}_a^m + \overline{E_a'{\bf v}'} 
\end{equation}
\begin{equation}
    \varepsilon_p^t = \overline{\varepsilon_p} - \varepsilon_p^m  
\end{equation}
Using the fact that ${\bf J}_a^m = -\Upsilon_m \kappa \nabla \overline{\rho}$, $\varepsilon_p^m = \kappa \nabla \overline{\rho} \cdot \nabla \Upsilon_m$, and 
$\overline{\varepsilon_p} = \kappa \nabla \overline{\rho} \cdot \nabla \overline{\Upsilon} + \kappa \overline{\nabla \rho'\cdot \nabla \Upsilon'}$, 
these can be more explicitly written as 
\begin{equation}
    {\bf J}_a^t = - \overline{\kappa \Upsilon' \nabla \rho'} - (\overline{\Upsilon} - \Upsilon_m) \kappa \nabla \overline{\rho} + \overline{E_a'{\bf v}'} 
\end{equation}
\begin{equation}
  \varepsilon_p^t 
  = \overline{\varepsilon_p} - \varepsilon_p^m 
  = \kappa \nabla \overline{\rho} \cdot 
  \nabla (\overline{\Upsilon} - \Upsilon^m ) 
  + \kappa \overline{\nabla \rho'\cdot 
  \nabla \Upsilon'} 
\end{equation}
These expressions show that the turbulent diffusive flux of APE and APE dissipation rate both depend on the mean quantity $\overline{\Upsilon}-\Upsilon_m$, whose leading order expression can be showed to be 
\begin{equation}
    \overline{\Upsilon} - \Upsilon_m 
    = \frac{g}{\rho_{\star}} 
    \left ( \overline{z_0(\rho)} - z_0 (\overline{\rho} )
    \right ) 
    \approx \frac{g}{\rho_{\star}} 
    \frac{\partial^2 z_0}{\partial \rho^2} 
    (\overline{\rho}) \frac{\overline{\rho'^2}}{2} 
\end{equation}
The above expressions also depends on $\Upsilon'$, which can be shown to be given at leading order
\begin{equation}
    \Upsilon' = \frac{g}{\rho_{\star}} 
    \left ( \overline{z_0(\rho)} - z_0(\rho) \right ) 
    \approx - 
    \frac{g}{\rho_{\star}} 
    \frac{\partial z_0}{\partial \rho}(\overline{\rho}) 
    \rho' 
    + \frac{g}{\rho_{\star}} 
    \frac{\partial^2 z_0}{\partial \rho^2} (\overline{\rho})
    \left ( \frac{\overline{\rho'^2}}{2} 
    - \frac{\rho'^2}{2} \right ) 
\end{equation}
As a result, it follows that 
\begin{equation}
    \kappa \nabla \overline{\rho} \cdot 
    \nabla ( \overline{\Upsilon} - \Upsilon_m) 
    \approx \frac{g}{\rho_{\star}} 
    \frac{\partial^2 z_0}{\partial \rho^2}(\overline{\rho} ) 
    \kappa \nabla \overline{\rho} \cdot 
    \nabla \frac{\overline{\rho'^2}}{2} 
    + \frac{g}{\rho_{\star}} \frac{\overline{\rho'^2}}{2} 
    \frac{\partial^3 z_0}{\partial \rho^3} (\overline{\rho}) 
    \kappa |\nabla \overline{\rho}|^2  
\end{equation}
\begin{equation}
    \kappa \overline{\nabla \rho' \cdot \nabla \Upsilon'} 
    \approx - \frac{g}{\rho_{\star}} 
    \frac{\partial z_0}{\partial \rho}(\overline{\rho}) 
    \kappa \overline{|\nabla \rho'|^2} 
    - \frac{g}{\rho_{\star}} 
    \frac{\partial^2 z_0}{\partial \rho^2} 
    (\overline{\rho}) \kappa \nabla \overline{\rho} \cdot 
    \nabla \frac{\overline{\rho'^2}}{2}  
\end{equation}
Summing up these two results, retaining only the terms up to second order in density perturbation, gets rid of the term proportional to $\partial^2 z_0/\partial \rho^2$ and yields the following equation for the turbulent APE dissipation rate
\par \medskip 
\fbox{
\begin{minipage}{8cm} 
\begin{equation}
     \varepsilon_p^t \approx - \frac{g}{\rho_{\star}} 
      \frac{\partial z_0}{\partial \rho}(\overline{\rho}) 
      \kappa \overline{|\nabla \rho'|^2} 
     {\color{black}+} \frac{g }{\rho_{\star}} 
     \frac{\overline{\rho'^2}}{2} 
     \frac{\partial^3 z_0}{\partial \rho^3} 
     (\overline{\rho}) \kappa  |\nabla \overline{\rho}|^2 
      \label{approximate_eddy_ape_dissipation}
\end{equation}
\end{minipage}
}
\par \medskip \noindent 

Of the two terms appearing in the right-hand side of (\ref{approximate_eddy_ape_dissipation}), only the first one can be ascertained to be non-negative and directly comparable to the expression for the APE dissipation rate proportional to the turbulent dissipation rate of 
density variance previously derived by \citet{Oakey1982} and 
\citet{Gargett1984b}. The second term, however, can be of any sign, but is expected to be much smaller than the first term in general and therefore unlikely to be important in practice, although this remains to be checked more systematically in direct numerical simulations of turbulent stratified mixing. This is left for future work. \citet{Scotti2014} only retained the first term
in the right-hand side of (\ref{approximate_eddy_ape_dissipation}) in their
paper.

\section{Application to energetically consistent numerical ocean modelling}
\label{constraint_on_mixing} 

Our framework provides a rigorous theoretical foundation for linking parameterised energy transfers to observable KE and APE dissipation rates. In numerical ocean models, the main energy transfers of interest are those associated with the ocean meso-scale, which are responsible for transferring mean APE to turbulent or eddy kinetic energy (which is subsequently dissipated through irreversible viscous processes), and those associated with small-scale turbulent stratified mixing, which transfer turbulent kinetic energy into turbulent APE, later dissipated by irreversible diffusive processes, giving rise to turbulent diapycnal mixing. The following aims to illustrate these ideas.

\subsection{Energetically consistent modelling} 

Parameterisations of subgridscale processes control the energy transfers between the resolved and unresolved scales of motion implicated in the turbulent forward KE and APE energy cascades. Upon reaching molecular scales, mechanical energy can be dissipated quasi-adiabatically as KE at the viscous dissipation rate $\varepsilon_k^t$, or diabatically as APE at the diffusive dissipation rate $\varepsilon_p^t$. \textcolor{black}{In most numerical ocean models, subgrid-scale process parameterisations have traditionally been developed without explicit consideration of their implications for the KE and APE cascades or for the KE and APE dissipation rates. However, this changed when \citet{Munk1998} noted that the intensity of mixing depends on the intensity of the mechanical sources of stirring. In particular, they raised the issue of whether tuning the value of the turbulent diapycnal or vertical mixing coefficients $K_{\rho}$ or $K_v$ to minimise mismatch with observed temperature and salinity fields was necessarily consistent with our understanding of how turbulent mixing is mechanically sustained. To that end, they considered the gravitational potential energy (GPE) budget, assuming a balance between the rate of GPE increase due to mixing at the rate $K_v N^2$ and the rate of GPE loss due to high-latitude cooling (neglecting the role of nonlinearities in the equation of state). To link $K_v$ to the mechanical sources of stirring, they used the relation $K_v N^2 = \Gamma \varepsilon_K$ with $\Gamma = 0.2$, and balanced the total viscous dissipation rate with the power input due to winds and tides, dismissing the role of surface buoyancy fluxes (see \citet{Tailleux2010} for a discussion). The work of \citet{Munk1998} proved very influential, prompting the development of `energetically consistent modelling' pioneered by Carsten Eden and his group \citep{Eden2014,Eden2015,Eden2016}, whose aim is to link the intensity of mixing to the sources of stirring by exploiting empirical knowledge about the dissipation ratio $\Gamma = \varepsilon_p/\varepsilon_k$. In our view, this aim can be more generally defined as seeking to exploit existing and developing knowledge about the inter-relations between the KE and APE budgets, for instance by developing and using a turbulent APE budget in addition to the widely used turbulent KE budget.}

In practice, the prevailing approach has primarily relied on using a turbulent kinetic energy (TKE) equation to predict the turbulent viscous dissipation $\varepsilon_k^t$, from which the APE dissipation rate $\varepsilon_p^t$ and turbulent vertical mixing coefficient $K_{\rho} = \varepsilon_p^t/N^2$ can be inferred, provided that the value of the dissipation ratio $\Gamma$ can also be predicted in some way. For a review of our current understanding of $\Gamma$, see \cite{Gregg2021}. An alternative, yet to be developed in oceanography, would be to predict $\varepsilon_p^t$ directly from a turbulent APE equation, as proposed in the context of atmospheric boundary layer research by Zilitinkevich and collaborators (e.g., \citet{Zilitinkevich2013}). An important advantage of the eddy APE budget is that it constrains the full turbulent density flux $\overline{\rho'{\bf v}'}$, whereas the TKE budget depends only on the vertical component $\overline{\rho'w'}$, \textcolor{black}{which is why we think it should receive more attention.}

To analyse across-scale energy transfers, a filtering approach is needed to isolate the different scales of interest. In the context of large-scale ocean modelling, \citet{Eden2015} proposed that the subgrid-scale energy should be divided into subreservoirs for meso-scale eddies, internal gravity waves, and turbulent incoherent motions. Here, we only consider a subdivision of energy into mean (resolved) and eddy (unresolved) scales due to the inherent limitations of standard Reynolds averaging. With this in mind, we return to the eddy APE budget (\ref{ape_density_evolution_first_form}), which we rewrite as
\begin{equation}
    \frac{\partial E_a^t}{\partial t} 
    + \nabla \cdot \left ( \overline{\bf v} E_a^t 
    + {\bf J}_a^t \right )  = C(E_k^t,E_a^t) + 
    C(E_a^m,E_a^t) - \varepsilon_p^t
    \label{eddy_APE_budget_new} 
\end{equation}
where $C(E_k^t,E_a^t)$ and $C(E_a^m,E_a^t)$ are the conversions of turbulent kinetic energy and mean APE into eddy APE, respectively, with
\begin{equation}
    C(E_k^t, E_a^t) = - \overline{b_{\ell}'w'} = 
    \frac{g}{\rho_{\star}} \overline{\rho'w'} ,
    \label{eke_to_eape}
\end{equation}
\begin{equation} 
      C(E_a^m,E_a^t) = -\overline{\rho'{\bf v}'} \cdot \nabla \Upsilon_m  =     - \frac{g}{\rho_{\star}} \overline{\rho'w'} 
    + \frac{g}{\rho_{\star}} \frac{\partial z_0}{\partial \rho} 
    (\overline{\rho}) \overline{\rho'{\bf v}'} \cdot 
    \nabla \overline{\rho} 
      \label{mape_to_eape} 
\end{equation}
Under the classical assumptions of stationarity and homogeneity, the terms on the left-hand side of (\ref{eddy_APE_budget_new}) can be neglected, and the eddy APE budget reduces to a balance between production of eddy APE due to conversions with mean APE and eddy KE, and turbulent APE dissipation, viz.,
\begin{equation}
    C(E_k^t, E_a^t) + C(E_a^m, E_a^t) \approx 
    \varepsilon_p^t 
\end{equation}
which, from (\ref{eke_to_eape}) and (\ref{mape_to_eape}), may be written as
\begin{equation}
    \frac{g}{\rho_{\star}} 
    \frac{\partial z_0}{\partial \rho} (\overline{\rho}) 
    \overline{\rho'{\bf v}'} \cdot \nabla \overline{\rho} 
    \approx \varepsilon_p^t .
    \label{ape_budget_balance} 
\end{equation}
Eq. (\ref{ape_budget_balance}) can be understood as a constraint on the diapycnal component of the turbulent density flux across the mean isopycnal surfaces $\overline{\rho} = {\rm constant}$, controlled by the turbulent APE dissipation rate $\varepsilon_p^t$. When the resolved flow pertains to large scales, the turbulent density flux is generally assumed to contain at least two components:
\begin{equation}
    \overline{\rho'{\bf v}'} = \overline{\rho'{\bf v}'}_{meso} + 
    \overline{\rho'{\bf v}'}_{small} 
\end{equation}
pertaining to the effects of meso-scale eddies and small-scale turbulent mixing, respectively, and are taken to be perpendicular and parallel to $\nabla \overline{\rho}$ according to
\begin{equation}
    \overline{\rho'{\bf v}'}_{meso} = \boldsymbol{\Psi} \times \nabla 
    \overline{\rho}  , \qquad 
    \overline{\rho'{\bf v}'}_{mix} = - K_{\rho} \nabla \overline{\rho} . 
    \label{turbulent_density_parameterisation} 
\end{equation}
(e.g., \citet{Griffies1998,Griffies1998b}). Of particular interest is the vertical component
\begin{equation}
    \overline{\rho'w'}_{mix} = - K_{\rho} 
    \frac{\partial \overline{\rho}}{\partial z}
\end{equation}
which will be later contrasted with the vertical component of the skew diffusive flux.

\subsection{Energetics of downgradient diffusion}

In the literature, the turbulent diffusivity $K_{\rho}$ entering the turbulent mixing parameterisation for the diffusive part of the turbulent density flux is traditionally predicted by
\begin{equation}
    K_{\rho} \approx  
  \frac{\varepsilon_p^t}{\overline{N}^2} 
  = \frac{\Gamma \varepsilon_k^t}{\overline{N}^2} 
  \label{osborn_cox_model}
\end{equation}
(e.g., \citet{Lindborg2008}), where $\Gamma$ is the dissipation ratio \cite{Oakey1982}. However, according to the eddy APE budget (\ref{turbulent_density_parameterisation}), a more accurate expression is
\begin{equation}
    - \frac{g}{\rho_{\star}} 
    \frac{\partial z_0}{\partial \rho} (\overline{\rho}) 
    K_{\rho} |\nabla \overline{\rho} |^2 
    \approx \varepsilon_p^t 
\end{equation}
which may be rearranged as
\par \medskip 
\fbox{
\begin{minipage}{8cm}
\begin{equation}
    K_{\rho} \approx \frac{1}{\Lambda (1+|{\bf S}|^2)} \frac{\varepsilon_p^t}{\overline{N}^2} \ne \frac{\varepsilon_p^t}{\overline{N}^2} 
    \label{new_krho} 
\end{equation}
\end{minipage}
}
\par \medskip \noindent 
where $\Lambda$ and ${\bf S}$ are defined as before by (\ref{n2_density_slope}). As stated earlier, Eq. (\ref{new_krho}) shows that the standard expression (\ref{osborn_cox_model}) implicitly depends on two assumptions that are rarely, if ever, acknowledged: 1) that the local stratification as measured by $\overline{N}^2$ is approximately equal to $N_0^2$; and 2) the small slope approximation $|{\bf S}|\ll 1$, often made in the context of rotated Redi diffusion \citep{Redi1982}, for instance. 
Moreover, if we neglect the higher order terms in the definition of the eddy APE dissipation rate so that
\begin{equation}
     \varepsilon_p^t \approx - \frac{g}{\rho_{\star}} 
     \frac{\partial z_0}{\partial \rho} (\overline{\rho} ) 
     \kappa \overline{|\nabla \rho'|^2} 
\end{equation}
it is easily verified that the eddy APE budget also implies
\par \medskip 
\fbox{
\begin{minipage}{8cm} 
\begin{equation}
     K_{\rho} |\nabla \overline{\rho} |^2 \approx \kappa \overline{|\nabla \rho'|^2}.
     \label{osborn_cox_model_variance} 
\end{equation}
\end{minipage}
}
\par \medskip \noindent 
Physically, Eq. (\ref{osborn_cox_model_variance}) states that the resolved dissipation of the mean density field must ultimately be balanced by the dissipation of eddy density variance at molecular scales. Note, however, that when approached from the eddy APE budget viewpoint, (\ref{osborn_cox_model_variance}) requires neglecting the higher order terms in the eddy APE dissipation rate. In the literature, (\ref{osborn_cox_model_variance}) is more commonly obtained by equating the production of density variance by the turbulent density flux with the dissipation of density variance, as in the Osborn-Cox model \citep{Osborn1972}. Note that Eq. (\ref{osborn_cox_model_variance}) cannot give rise to upgradient (negative) diffusion, while (\ref{new_krho}) can potentially allow it in some circumstances.

To synthesise the results, the energy conversions affected by the downgradient part of the turbulent density flux are therefore given by
\par \medskip 
\fbox{
\begin{minipage}{8cm} 
\begin{equation}
   \left \{ C(E_a^m,E_a^t) \right \}_{mix} 
    = \left ( (1+|{\bf S}|^2 ) \Lambda -1 \right )  K_{\rho} \overline{N}^2 
    \label{c_eam_eat_final}
\end{equation}
\begin{equation} 
   \left \{ C(E_k^t,E_a^t) \right \}_{mix} 
    = K_{\rho} \overline{N}^2 .
    \label{c_ekt_eat_final} 
\end{equation} 
\end{minipage}
} 
\par \bigskip \noindent 
In the literature, the mean to eddy APE conversion has been exclusively discussed in the context of the quasi-geostrophic (QG) approximation (e.g., \citet{vonStorch2012}), which is equivalent to assuming $\Lambda \approx 1$ in (\ref{c_eam_eat_final}), in which case it reduces to
\begin{equation}
    \{ C(E_a^m,E_a^t) \}_{qg,mix} \approx 
    |{\bf S}|^2 K_{\rho} \overline{N}^2 \ge 0 ,
    \label{eam_to_eap_qg}
\end{equation}
and always acts as a downscale transfer of energy from mean APE to eddy APE. 
The exact finite-amplitude mean APE to eddy APE conversion may occasionally behave quite differently, as Eq. (\ref{c_eam_eat_final}) shows that the sign of the conversion is no longer necessarily non-negative. Energy transfer can be either upscale or downscale, depending on the relative values of $\Lambda$ and the slope parameter $|{\bf S}|$. 
\textcolor{black}{Because the value of $\Lambda$ is controlled by the distance from the Lorenz reference state, which itself depends on the steepness of isopycnal slopes, we assume that the condition for upscale transfer from eddy APE to mean APE imposes a constraint on $\Lambda$ (rather than on $|{\bf S}|$), namely
\begin{equation}
    \Lambda < \frac{1}{1+|{\bf S}|^2} .
    \label{condition_instability}
\end{equation}
}
In that case, downgradient diffusion can, at least in principle, backscatter unresolved energy into resolved energy, thus potentially acting as a source of instability for the resolved flow. Although the associated upscale energy transfer will be counteracted by the downscale energy transfer associated with skew diffusion, it is not necessarily obvious that this can actually suppress the diffusive instability, because downgradient diffusion and skew diffusion a priori operate differently.

\subsection{Energetics of skew diffusion and TKE budget} 

Physically, skew diffusion can be interpreted as an eddy-induced advection by subgrid-scale processes, as follows from the relation
\begin{equation} 
\nabla \cdot ( \boldsymbol{\Psi} \times \nabla \overline{\rho} ) = (\nabla \times 
\boldsymbol{\Psi}) \cdot \nabla \overline{\rho} =
{\bf v}_{eddy} \cdot \nabla \overline{\rho} .
\end{equation} 
As a result, the evolution equation for the mean density $\overline{\rho}$ may be written as
\begin{equation}
      \frac{\partial \overline{\rho}}{\partial t} + 
      {\bf v}_{res} \cdot \nabla \overline{\rho} = \nabla \cdot [ (K_{\rho} + \kappa )
      \nabla \overline{\rho} ]
\end{equation}
with ${\bf v}_{res} = \overline{\bf v}+ {\bf v}_{eddy}$ being the residual velocity and ${\bf v}_{eddy} = \nabla \times \boldsymbol{\Psi}$ the eddy-induced velocity. By construction, the eddy-induced velocity is divergenceless, $\nabla \cdot {\bf v}_{eddy} = 0$. In oceanography, the most commonly used parameterisation for $\boldsymbol{\Psi}$ is  
\begin{equation}
    \boldsymbol{\Psi} = \boldsymbol{\Psi}_{gm}  = 
    {\bf k} \times \kappa_{gm} {\bf S} ,
    \label{gm_parameterisation} 
\end{equation}
as originally proposed by \citet{Gent1990} and \citet{Gent1995}, where ${\bf S}$ is the slope vector previously introduced. For the GM parameterisation, the skew-diffusion part of the turbulent density flux becomes
\begin{equation} 
   \overline{\rho'{\bf v}'}_{skew} = ( {\bf k} \times \kappa_{gm} {\bf S} ) 
   \times \nabla \overline{\rho} 
   = \kappa_{gm} \frac{\partial \overline{\rho}}{\partial z} {\bf S} 
   + \kappa_{gm} |{\bf S}|^2 \frac{\partial \overline{\rho}}{\partial z} {\bf k} .
\end{equation} 
Of particular interest is the vertical component, given by
\begin{equation}
    \overline{\rho'w'}_{skew} = \kappa_{gm} |{\bf S}|^2 
    \frac{\partial \overline{\rho}}{\partial z}  
    \label{skew_vertical_flux} ,
\end{equation}
which \textcolor{black}{acts as an `upgradient' flux}. 

The skew-diffusive part of the turbulent density flux is perpendicular to the mean density gradient $\nabla \overline{\rho}$, and therefore does not have a net contribution to the eddy APE budget. However, it is associated with net energy conversions between mean and eddy APE, as well as between eddy APE and eddy KE. From (\ref{eke_to_eape}) and (\ref{skew_vertical_flux}), it is easily verified that
\par \medskip 
\fbox{
\begin{minipage}{8cm} 
\begin{equation}
\begin{split} 
    \left \{ C(E_a^m,E_a^t) \right \}_{skew}  
    = & - \overline{\rho'{\bf v}'}_{skew} \cdot \nabla \Upsilon_m \\
    = &  - \frac{g}{\rho_{\star}} \overline{\rho'w'}_{skew} 
    = \kappa_{gm} |{\bf S}|^2 \overline{N}^2 
    \label{gm_eam_eat_conversion}
 \end{split} 
\end{equation}
\begin{equation}
  \left \{ C(E_k^t,E_a^t) \right \}_{skew} 
  = \frac{g}{\rho_{\star}} \overline{\rho'w'}_{skew} 
   = - \kappa_{gm} 
   |{\bf S}|^2 \overline{N}^2 
   \label{gm_ekt_to_eat_conversion}
\end{equation}
\end{minipage}
}
\par \medskip \noindent 
Eq. (\ref{gm_eam_eat_conversion}) confirms that skew diffusion acts as a net sink of mean APE whose magnitude is proportional to the slope squared $|{\bf S}|^2$ and mean squared buoyancy frequency $\overline{N}^2$ \textcolor{black}{e.g., \citet{Griffies2004}}.

\subsection{Remarks on the eddy (turbulent) kinetic energy budget}

Since skew diffusion associated with the meso-scale eddy parameterisation acts as a net source of eddy KE rather than of eddy APE, it is useful to conclude this section with some comments on the eddy KE budget and how the KE and APE cascades may help constrain mixing parameterisations. Under stationary and homogeneous conditions, the main sources of eddy KE are:
\begin{enumerate} 
\item the downscale energy transfer associated with the mixing of momentum, primarily contributed by the vertical shear,
\begin{equation}
     C(E_k^m,E_k^t) \approx A_v \left ( \frac{\partial \overline{\bf u}}{\partial z}
     \right )^2 ,
\end{equation}
\item the downscale transfer associated with the meso-scale eddy parameterisation, which controls the conversion $C(E_a^m,E_k^t)$; 
\item the loss of energy associated with diapycnal mixing, which controls $C(E_k^t,E_a^t)$; 
\item the viscous dissipation rate, dissipating $E_K^t$ into background potential energy or `heat'. 
\end{enumerate} 
Summing up all these contributions leads to the following balance:
\begin{equation}
    A_v \left ( \frac{\partial \overline{\bf u}}{\partial z} \right )^2 
    + \kappa_{gm} |{\bf S}|^2 \overline{N}^2 
    - K_{\rho} \overline{N}^2 \approx \varepsilon_k^t .
\end{equation}
To show the dependence of this balance on the Richardson number, it is useful to divide this relation by $\overline{N}^2$, yielding
\begin{equation}
    A_v R_i^{-1} + \kappa_{gm} |{\bf S}|^2 \approx K_{\rho} 
    + \frac{\varepsilon_k^t}{\overline{N}^2} 
\end{equation}
where
\begin{equation}
     R_i = \overline{N}^2 \left ( \frac{\partial \overline{\bf u}}{\partial z} 
     \right )^{-2} 
\end{equation}
is the Richardson number. This relation shows that the different turbulent mixing parameters $K_v$, $\kappa_{gm}$, and $K_{\rho}$ are not independent from each other. Notably, by using the relation $K_{\rho} \approx \Gamma \varepsilon_k^t/\overline{N}^2$, the above relation can be written as
\begin{equation}
    A_v R_i^{-1} + \kappa_{gm} |{\bf S}|^2 
    \approx \frac{(\Gamma+1) \varepsilon_k^t}{\overline{N}^2} 
\end{equation}
and can be interpreted as a diagnostic energy balance potentially useful for predicting the turbulent kinetic energy dissipation rate, which is a crucial element of energetically consistent meso-scale eddy parameterisations such as in the GEOMETRIC framework (e.g., \citet{Marshall2012,Mak2018,Torres2023}) or in \citet{Jansen2015,Jansen2019}, for instance.
In regions where the vertical shear can be predicted by thermal wind balance,
\begin{equation}
    \frac{\partial \overline{{\bf u}}}{\partial z} \approx -\frac{g}{\rho_{\star} f} {\bf k} \times \nabla \overline{\rho}, 
\end{equation} 
the TKE budget may also be written as
\begin{equation}
     \left ( A_v \frac{\overline{N}^2}{f^2} 
     + \kappa_{gm} \right ) |{\bf S}|^2
     \overline{N}^2
     \approx (\Gamma+1) \varepsilon_k^t .
     \label{new_relation}
\end{equation}
Eq. (\ref{new_relation}) shows that in this regime, the APE cascade converting mean APE into eddy APE behaves analogously to the KE cascade converting mean KE into eddy KE, similarly to \citet{Lindborg2006}'s relations (\ref{lindborg_relations}), provided that 
\begin{equation}
    \kappa_{gm} \approx \Gamma A_v 
    \frac{\overline{N}^2}{f^2} 
\end{equation}
which appears to be closely related to the case discussed by \citet{Greatbatch1990}. 
Determining to what extent the present results can inform further developments of such parameterisations is beyond the scope of this paper and will be addressed in a subsequent study.

\section{Summary and discussion} 
\label{summary_and_discussion} 

In this study, we have revisited the mean/eddy decomposition theory for local Available Potential Energy (APE) density, focusing on its application to the characterization of meso-scale eddies, small-scale irreversible mixing, and their interaction with the large-scale circulation. While irreversible mixing and eddy features occur at scales much smaller than the planetary-scale circulation, these processes play a crucial role in the global energy budget, necessitating careful consideration to ensure that forcing and dissipation terms are computed in a mutually consistent manner. Our approach addresses the consistency issues associated with the definition of localized forms of local APE density, which are often used to describe the energetics of these processes. Our analysis demonstrates that a consistent formulation should be based on partitioning the local APE density into mean and eddy components. The resulting eddy APE density measures forces against the local mean density profile, aligning with physical intuition. This formulation differs from heuristic localized forms of APE density, which ignore the dependence on the Lorenz reference state. Our exact eddy APE density retains some dependence on the Lorenz reference state, differing from the heuristic form by a factor of $\Lambda = \overline{N}^2/N_0^2$. Previous studies indicate that while fluid parcels are very close to their reference position in most of the ocean interior, where $\Lambda \approx 1$, this is generally not the case in the polar regions or the Gulf Stream area (e.g., \citet{Saenz2015,Tailleux2016b,Tailleux2013}) where our
results should be of most practical use.

\smallskip 

Our approach is physically and conceptually simpler than that of \citet{Scotti2014}, while being mathematically equivalent, as discussed in detail by \citet{Tailleux2024}. The exact eddy APE budget is easier to interpret, as its net conversion with mean APE and eddy KE depends solely on the diapycnal component of the turbulent density flux, in contrast to the quasi-geostrophic (QG) approximated version. This formulation is valid for arbitrarily large departures from the Lorenz reference state, characterized by the parameters $\Lambda$ and the norm of the slope vector $|{\bf S}|$. A fundamental difference between the two frameworks concerns the mean to eddy APE conversion, which can only be positive in the QG approximation, but which can potentially be also negative in the exact case. The exact eddy APE density behaves similarly to the heuristic localized form, despite being defined by different mathematical expressions. Both forms agree that the buoyancy involved in local turbulent stirring/mixing is the buoyancy anomaly defined relative to the mean density field, but they differ in how they define the displacement, $\zeta' = z -z_m$ versus $\zeta' = z_0(\rho) - z_0(\overline{\rho})$. This has important consequences for the study of meso-scale eddy APE and turbulent stratified mixing where the two differ significantly, notably regarding the theory for mixing efficiency and the determination of turbulent vertical mixing diffusivity, as is expected to be the case in the polar regions, for instance.

\smallskip 

Our approach represents a significant step towards a more realistic and accurate treatment of the APE budget, which is crucial for developing energetically consistent parameterizations and numerical ocean models. Importantly, such progress can only be achieved by the local theory of APE, highlighting the limitations of \citet{Winters1995}'s global APE framework. To fully assess the implications for estimating meso-scale eddy APE and the study of turbulent stratified mixing in the oceans, future work will involve reformulating the present framework using a Large Eddy Simulation (LES) spatial filter instead of Reynolds averaging and accounting for the nonlinearities of the equation of state for seawater. Preliminary results suggest that these nonlinearities can occasionally cause a loss of convexity for the local APE density, potentially corresponding to thermobarically unstable situations. The exact mean/eddy decomposition of the APE density, while simple, provides valuable insights into the turbulent APE cascade. Understanding how to rigorously apply this decomposition to finite-amplitude APE density is the first step towards a true multi-scale analysis, of which spectral analysis is one example. Further research is needed to develop a joint multi-scale analysis of APE density and kinetic energy (KE) that can provide more insights into how information about the Lorenz reference state affects different scales.

\smallskip 
In conclusion, our study has clarified the relationship between heuristic and exact forms of localized eddy APE density and has demonstrated the importance of considering the dependence on the Lorenz reference state for a consistent treatment of the APE budget. This work lays the foundation for future research into the development of energetically consistent ocean mixing parameterizations and the multi-scale analysis of APE and KE in turbulent stratified flows.

\appendix

\bibliographystyle{elsarticle-harv} 





\end{document}